\documentclass[extra,referee]{gji}

\usepackage{color}
\usepackage{timet}
\usepackage{graphicx}

\title[A new 1D velocity model of western Rift of Corinth]
  {A new 1D $V_p$ and $V_s$ velocity model of the western Rift of Corinth, Greece,
   using a fully non-linear tomography algorithm}

\author[M. Noble et al.]
  {
  Mark Noble$^1$\thanks{Corresponding author},
  Alexandrine Gesret$^1$,
  H\'el\`ene Lyon-Caen$^2$,
  Anne Deschamps$^3$\\
  $^1$ MINES Paris -- PSL University, Centre de G\'eosciences, Fontainebleau, France\\
  $^2$ Laboratoire de G\'eologie, CNRS, \'Ecole Normale Supérieure, PSL University, Paris, France\\
  $^3$ Universit\'e Côte d'Azur, CNRS, Observatoire de la C\^ote d'Azur, IRD, Géoazur, Nice, France
  }
\date{2022/01/19}
\pagerange{\pageref{firstpage}--\pageref{lastpage}}
\volume{200}
\pubyear{2022}

\begin{document}

\label{firstpage}

\maketitle

\begin{summary}
The objective of this study is to propose a new updated accurate 1-D $V_p$ and $V_s$ velocity model of the western Rift of Corinth for precise absolute earthquake locations.
The methodology used to obtain this new model associates two fully non-linear algorithms, a complete exploration of the $V_p$ and $V_s$ model space combined with a grid search method for earthquake locations to minimize the P and S arrival time residuals.
We calculated the misfit function for approximately $2.10^6$ velocity models.
For the best model (minimum misfit) we observed a global decrease of 30 $\%$ for both the P and S residuals with respect to the commonly used velocity model.
The main features of this new model compared to other studies is a significant
decrease of $V_p$ and $V_s$ velocities from surface down to a depth of approximately 3 km and a variable ${V_p} /{V_s}$ ratio decreasing from 1.86 at the surface to 1.78 at a depth of 8 km.
The major influence on the locations of events is a global decrease of focal depths that range from a few hundred metres for deep events to more than one kilometre for shallow earthquakes.

\end{summary}

\begin{keywords}
Western Rift of Corinth, 1D velocity model, uppermost crust, $V_{p}$, $V_{s}$, $V_{p}/V_{s}$ ratio
\end{keywords}

\section{Introduction}
\begin{figure}
\centering
\includegraphics[width=12cm]{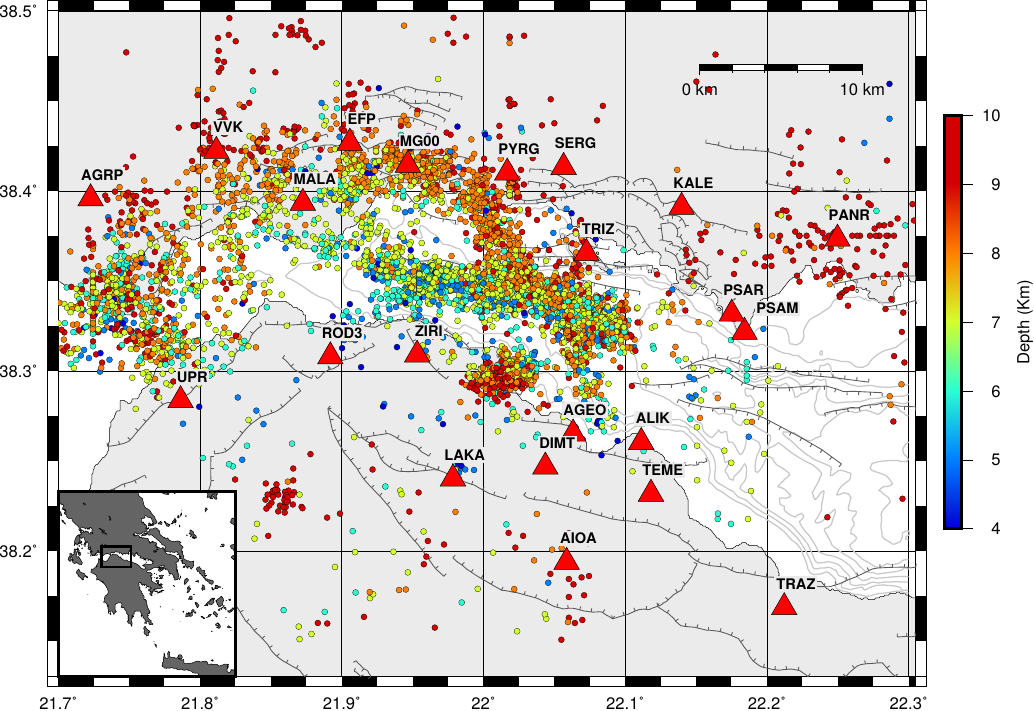}

\caption{Map of the Western Rift of Corinth, the location in central Greece is shown in the inset map.
Seismological stations of the Corinth Rift Laboratory (CRL) network are depicted by red triangles. 
To illustrate the strong seismicity recorded by the network (obtained by automatic phase picking and location) and its spatial distribution across the Western Rift of Corinth, epicentres of one year of seismicity (2017) are depicted circles, colour of events indicate focal depth.
In 2017 more than 36 000 events where recorded, only events with a minimum of 11 $P$ and 7 $S$ first-arrival time readings, a maximum horizontal location error of 1 km and a maximum vertical error of 2 km and a maximum rms of 0.45 s, are displayed (7400 events). For description of the major faults, see \cite{duverger_dynamics_2018}.}
\label{Fig1}
\end{figure}

The rift of Corinth (central Greece) is an active asymmetric graben extending east-west over approximately 105 km with a width of 10--30 km.
It is bounded by continental Greece to the north and northern Peloponnesus to the south  (Fig. \ref{Fig1}).
The coast shape defines an approximately $100^o$ N-oriented structure, which is narrow and shallow on the western side while it is wider and deeper on the eastern side.
During the last 30 years, many geological and geodetic studies have been carried out to try and understand the tectonic evolution of the Rift of Corinth.
This rift is characterized by a set normal faults, the major ones cropping out on the southern coast \citep[e.g.,][]{armijo_quaternary_1996} dipping north at $50^o - 70^o$.
This south coast is uplifting of several millimetres per year \citep[e.g.,][]{vita-finzi_evaluating_1993,armijo_quaternary_1996}. 
Offshore, in the middle of the gulf active faults have also been identified \citep[e.g.,][]{stefatos_seismic_2002,moretti_gulf_2003,bell_evolution_2008, beckers_active_2015}.

The northern border of the rift is characterized by minor antithetic south-dipping faults \citep[e.g.,][]{ori_geologic_1989, doutsos_geometry_1992, armijo_quaternary_1996, sorel_pleistocene_2000}.
The high interest for this active continental rift arises from the very strong strain rate reaching $10^{-5}$ $s^{-1}$ (1 cm/yr extension over 10 km) in the western rift of Corinth \citep{briole_active_2000, briole_gps_2021} resulting into a very strong seismic activity.
In average more than 10 000 events with magnitude ranging between about 1 and 4 are recorded every year and most of this seismicity is located between 6 and 8 km in depth.

Several instrumental and historical earthquakes of magnitude larger than 5.5 have been reported \citep{jackson_seismicity_1982, ambraseys_seismicity_1990, hatzfeld_microseismicity_2000, bernard_ms_1997}.
The seismicity  is also characterized by a significant background microseismic activity with numerous swarms clustered in time and space \citep[e.g.,][]{rigo_microseismic_1996, pacchiani_geometry_2010, sokos_january_2012, kapetanidis_2013_2015, duverger_20032004_2015, de_barros_imbricated_2020, kaviris_western_2021}.
To locate all these earthquakes most authors used the velocity model of \cite{rigo_microseismic_1996}.
\cite{rigo_microseismic_1996} used the data of the temporary seismological network that was operated for 2 months in 1991 \citep{le_meur_seismic_1997} and by following a trial-and-error approach, defined a multilayer 1-D $P$ velocity model with a constant $V_{p}/V_{s}$ ratio using the HYPO71 algorithm \citep{lee_hypo71_1975}.

Quite a few authors used this 1-D velocity profile proposed by \cite{rigo_microseismic_1996} as a starting model for three-dimensional iterative linearized delay traveltime tomography \citep[e.g.,][]{le_meur_seismic_1997, latorre_new_2004, gautier_new_2006}.
Very few attempts using fully non-linear algorithms have been made to refine or update the model proposed by \cite{rigo_microseismic_1996}.
\cite{jansky_estimation_2007} selected 55 events and used a nearest neighbourhood algorithm to propose a new $V_p$ profile.
\cite{giannopoulos_ambient_2017} performed an ambient noise tomography (Rayleigh waves) resulting in a $V_s$ profile down to a depth of 2 to 4 km. \cite{novotny_northwestern_2001} inferred $V_s$ model using mainly Love waves from distant earthquakes.

None of the two $V_s$ velocity profiles just mentioned are in agreement with the constant $V_{p}/V_{s}$ ratio of \cite{rigo_microseismic_1996}.
The objective of this study is to use the high-quality events recorded by Corinth Rift Laboratory permanent network (CRL, https://nfo.crlab.eu/) installed from the year 2000 onwards around the Western Rift of Corinth to invert for a new accurate minimum 1-D $V_{p}$ and $V_{s}$ velocity model.
An accurate velocity model is a key element for high-precision, absolute earthquake locations.
After reviewing the main previous velocity models of the Western Rift on Corinth proposed by different authors, we then describe the local seismic network and the data set used.
Then we present the non-linear inversion strategy and the new optimal minimum 1-D $V_{p}$ and $V_{s}$ velocity model.
We finally discuss the impact of this new velocity model on the locations of events.

\section{Velocity models from previous studies}
\begin{figure}
\centering
\includegraphics[width=11cm]{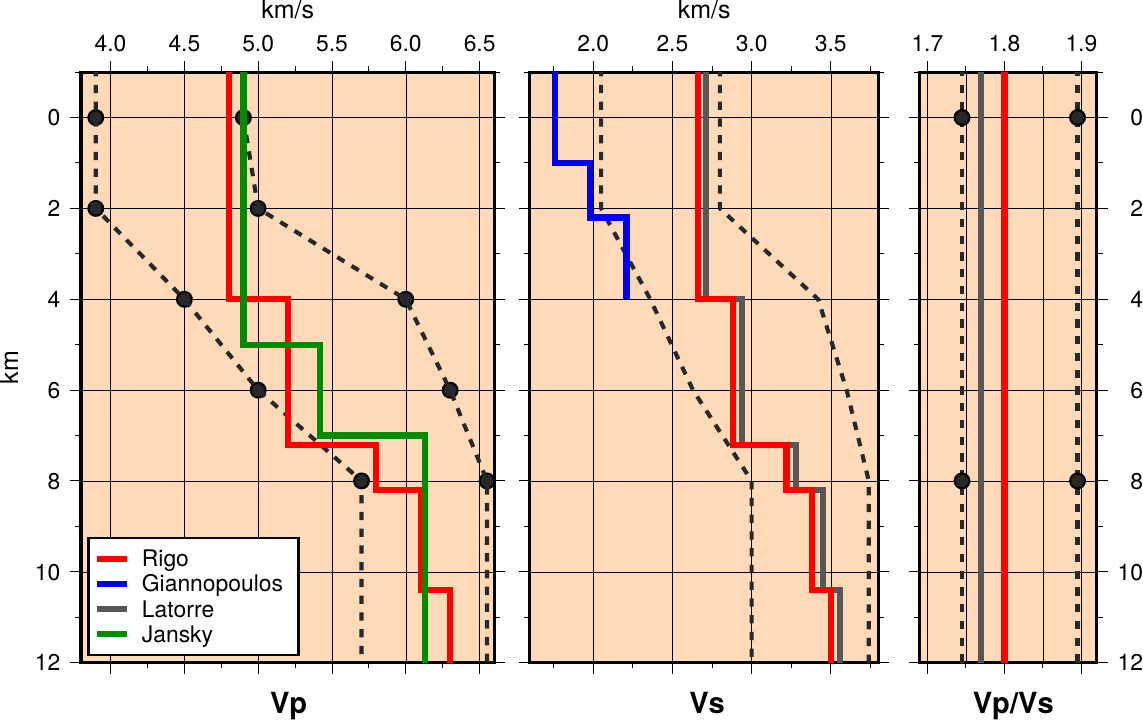}
\caption{
Left: $V_p$ velocity profiles of \cite{rigo_microseismic_1996} and \cite{jansky_estimation_2007}.
Middle:  $V_s$ velocity profiles of \cite{rigo_microseismic_1996}, \cite{latorre_new_2004} and \cite{giannopoulos_ambient_2017}.
Right $V_p/V_s$ ratios \cite{rigo_microseismic_1996}, \cite{latorre_new_2004}.
Dotted lines in all 3 graphs represent lower and upper bounds used to constrain our grid search tomography algorithm.
Black solid circles in left and right graph show the parametrization of models.
$P$ velocity profiles are defined by 5 nodes, $V_{p}/V_{s}$ profiles are defined by 2 nodes.
Between nodes $V_p$ and  $V_{p}/V_{s}$ are linearly interpolated.
Above first node at 0 km and below last node at 8 km, all values are set constant to nearest node.
Velocity model has been extended 1 km above sea-level to account for station elevations.
}
\label{Fig2}
\end{figure}

In this section we review the main $V_p$ and $V_s$ 1-D velocity profiles proposed in the literature concerning the uppermost crust of the western Gulf of Corinth. 
All the models presented in this section are displayed in Fig. \ref{Fig2} (first 12 kilometres in depth are displayed).

The first model was proposed by \cite{rigo_microseismic_1996} who used the data of the temporary seismological network that was operated for 2 months in 1991 \citep{le_meur_seismic_1997} and by following a trial-and-error approach, defined an optimal multilayer 1-D $P$ velocity model with a constant $V_{p}/V_{s}$ ratio using the HYPO71 algorithm \citep{lee_hypo71_1975}.
The model is composed of six homogeneous layers over a half-space with the Moho at a depth of 30 km.
The $V_{p}/V_{s}$ ratio determined from Wadati diagrams is constant and equal to 1.80. $P$ velocities range from 4.8 km/s at the surface up to a maximum of 6.5 km/s at depth with a very sharp velocity contrast between 7 and 8 km in depth.
The $V_s$ velocities follow exactly the same trend ranging from 2.66 km/s at the surface to 3.5 km/s at depth (\ref{Fig2}).

 \cite{jansky_estimation_2007} used 55 events whose epicentres are located in the middle of the Gulf to derive a new minimum 1-D velocity model of the uppermost crust.
The originality of this study is the use of two fully non-linear algorithms, a neighbourhood algorithm \citep{sambridge_geophysical_1999} combined with a grid search method for earthquake locations to minimize the $P$ and $S$ arrival time residuals.
However the number of observations per event was limited to 7 $P$ phases and 7 $S$ phases, all events were located in a narrow depth range between 6 and 8 km, limiting the accuracy of the inversion and constrained the authors to describe the velocity profile with only 3 layers.
The inversion did not produce any major update for the $S$ velocities.
To our knowledge this is the only study that attempted to used a full non-linear methodology to try and update the velocities of the uppermost crust.
This $V_p$ model is displayed in Fig. \ref{Fig2}.

\cite{latorre_new_2004}  used the 1-D P velocity profile proposed by \cite{rigo_microseismic_1996} as starting model for a three-dimensional iterative linearized delay traveltime tomography study of the data acquired with temporary network acquired in 1991, and found that a constant $V_{p}/V_{s}=1.77$ was more suitable.
This $V_{p}/V_{s}$ leads to higher $S$ velocities (Fig. \ref{Fig2}).

\cite{giannopoulos_ambient_2017} performed an ambient-noise interferometry tomography (Rayleigh waves) resulting in a 1D average $V_s$ profile that is probably well resolved down to a depth of 2 to 3  km, resulting in an average shear velocity around 2 km/s (Fig. \ref{Fig2}).
The derived shear velocities obtained from this ambient-noise interferometry study are quite lower than the aforementioned $V_s$ models derived from local traveltime measurements.

And finally we briefly mention the work of \cite{novotny_northwestern_2001} who inferred $V_s$ model using mainly Love waves from distant earthquakes. The obtained model displays even lower $S$ velocities (around 1.3 km/s) in the uppermost layers.

The discrepancy concerning the shear velocities between the different studies is a major argument to try and infer a new accurate model of the uppermost crust of the western Gulf of Corinth.
In the next section we present the local permanent seismic network and the selected data set used to try and better constrain the first few kilometres of the shallower crust.


\section{The CRL seismic network and selected data}

The first permanent stations of the CRL network have been installed in 2000 in the western part of the rift around the city of Aigion.
Between 2000 and 2009, the network covered a 30 km x 30 km area with 11 permanent stations, equipped with three-component seismometers.
In 2010, four new three-component velocimeters recording at 100 Hz have been installed further west to especially monitor the activity around the Psathopyrgos fault.
The CRLNET also includes seven stations equipped with broad-band seismometers operated by the University of Athens  and the University of Patras.
Nowadays, the CRL network covers a 60 km x 40 km area.
From 2010 onwards all stations are recording at the same sampling rate of 100 Hz.
The geographical position of the stations that we used for this study are displayed in Fig. (\ref{Fig1}).

The CRLNET catalog is built by replaying all continuous recording with \cite{helmholtz-centre_potsdam-gfz_german_research_centre_for_geosciences_seiscomp_2008}, using the scanloc module proposed by Gempa. This module uses a cluster search algorithm to associate P detected phases of a single earthquake and allows also for reliable S picks on horizontal components.

Preliminary location is done using HYPO71 software \citep{lee_hypo71_1975} and the 1-D velocity model of Rigo et al. (1996) with a constant $V_p/V_s$ ratio of 1.80.
When analysing the CRLnet catalogues, depending on the number of phases picked, the quality of picks, the average standard location errors that are output by HYPO71 can range from a few hundred metres to a few kilometres for events located inside the network.
For events outside the network location errors can reach 5 to 6 km and be even greater.
To illustrate the strong seismicity recorded by the network (including automatic phase picking and location) and its spatial distribution across the Western Gulf of Corinth, we displayed in Fig. \ref{Fig1} a subset of the 36 000 events recorded in 2017.
Most of this seismicity is located between 6 and 9 km in depth, no events can be found at depths between the surface and a depth of approximately 4 km.
For this subset the selection  criteria are the following: minimum of 11 $P$ and 7 $S$ first-arrival time readings, a maximum horizontal location error of 1 km and a maximum vertical error of 2 km and the final criterion is a maximum rms of 0.45 s, leading to a subset of 7400 events.

\begin{figure}
\centering
\includegraphics[width=13cm]{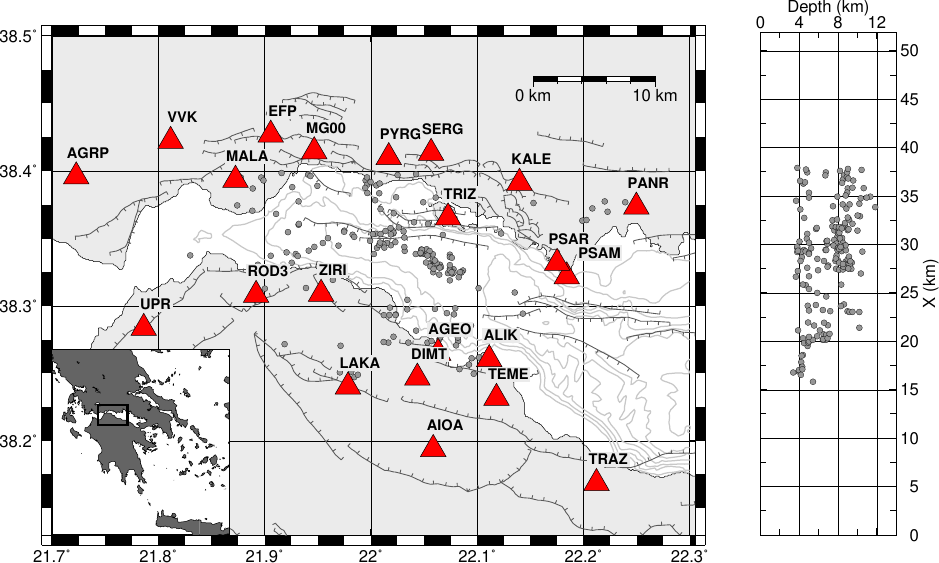}
\caption{Left: map of selected 193 events extracted from CRL catalog. Right: South - North vertical profile of same events.}
\label{Fig3}
\end{figure}

To try and better constrain the first few kilometres of the shallower crust of the western Rift of Corinth we are seeking high quality events recorded by the seismological network.
We used the CRLNET catalogues between 2015 and 2017.
Among the tens of thousands of events, we first selected from these catalogues only the events that were located inside the network to avoid large azimuthal gaps, with a minimum of 20 $P$ phases and a minimum of 15 $S$ phases.
The criterion we selected concerning the number of phases was acceptable for events located at depths greater than 5 km.
For events between the surface and 5 km, we had to release this criterion to 17 $P$ phases and 12 $S$ phases. In practice, we found no events at depths smaller than 3.8 km
In a second step only events with a maximum vertical error of 1.5 km and a maximum rms of 0.40 s were kept.
Finally to ensure a uniform distribution of events in depth, in average 20 to 30 events per km depth with the smallest rms were preserved. The final 193 selected events whose magnitude ranges from 0.9 to 3.1 are displayed in Fig. \ref{Fig3}. From the vertical section it can be seen that depth of events range from approximately 4 km down to 10 km, with only a few events between 10 and 12 km depth. As just mentioned previously no events with enough number of picked phases were found in the depth range from 0 to 4 km. Concerning the horizontal spatial distribution, events are well spread across the seismic network, meaning a wide range of epicentre-station distances that should contribute to better constrain velocities.
The final pre-processing step before the non linear tomography was the $P$ and $S$ arrival time manual picking. In average time picking accuracy was estimated in a range of 0.01 – 0.03 s for P and 0.02 – 0.07 s for S first-arrival times.

\section{Methodology}
\newcommand{\mb}[1]{\mathbf{#1}}
\newcommand{\ud}{\mathrm{d}}
\newcommand{\tobs}{{\mathbf{t}^{\textrm{\small obs}}}}
\newcommand{\tcal}{{\mathbf{t}^{\textrm{\small cal}}}}
\newcommand{\tobst}{{\tilde{\mathbf{t}}^{\textrm{\scriptsize obs}}}}
\newcommand{\tcalt}{{\tilde{\mathbf{t}}^{\textrm{\scriptsize cal}}}}
\newcommand{\tiobst}{{\tilde{t}^{\textrm{\scriptsize obs}}_{i}}}
\newcommand{\ticalt}{{\tilde{t}^{\textrm{\scriptsize cal}}_{i}}}
\newcommand{\tijobst}{{\tilde{t}^{\textrm{\scriptsize obs}}_{ji}}}
\newcommand{\tijcalt}{{\tilde{t}^{\textrm{\scriptsize cal}}_{ji}}}
When locating seismic events, the quality of the estimated hypocentre locations depend essentially on the velocity model accuracy (e.g. \cite{husen_understanding_2010}, \cite{gesret_propagation_2015}).
In order to obtain more reliable locations and the associated best velocity model, the traveltime data can be used to invert for both the hypocentres and the velocity model parameters by local earthquake tomography.
The most common  approach relies on iterative linearized inversion.
These algorithms aim to retrieve the maximum likelihood solution that corresponds to the minimum of the misfit function between observed and computed P and S traveltimes.
The major drawback of such local optimization methods is to give only a unique solution that can be inaccurate and is strongly dependent on the choice of the initial velocity model ($V_p$, $V_{p}/V_{s}$) and the initial event locations (coordinates and origin time).
Indeed as this problem is non-linear, its solution can be multimodal and the algorithm can be stuck in a local minimum that does not correspond to the true velocity model and hypocentre locations.
An alternative solution is to use a sampling approach of the a posteriori probability density function of event locations and velocity model that will give the global maximum of the misfit function.
\cite{piana_agostinetti_local_2015} applied such an approach but this 3D stochastic local earthquake tomography required weeks of computational time and they only inverted for a $V_p$ (accoustic approximation).
We propose here to use instead a sampling approach namely a grid search where the whole parameter space of both the 1D ($V_p$, $V_{p}/V_{s}$) velocity model and the earthquake locations (coordinates and origin time) is explored.
This should allow to avoid local minima of the misfit function in an efficient way.

When locating a single earthquake in a single velocity model, since the work of \cite{tarantola_inverse_1982}, the probabilistic Bayesian formulation is often chosen \citep[e.g., ][]{moser_hypocenter_1992, lomax_probabilistic_2000, husen_probabilistic_2003} as it allows to retrieve the global minimum of the misfit function even for a non-linear/multimodal solution.
This probabilistic approach allows to estimate the posterior probability density function (pdf) of event location given the observed picked traveltimes $\pi(x,y,z,t0|\tobs)$.
As shown by \cite{tarantola_inverse_1982}, if the prior information on the origin time $\pi(t0)$ is considered as uniform, the posterior pdf can be integrated analytically to obtain the marginal posterior pdf of the event spatial location $\pi(x,y,z|\tobs)$ which writes, 
\begin{equation}
\pi(x,y,z|\tobs) \propto \pi(x,y,z) \exp\left[- \frac{1}{2}(\tobst-\tcalt)^t\Sigma_s^{-1}(\tobst-\tcalt)\right].
\end{equation}
where $\tobst$ is the vector of observed arrival times $\tobs$ minus their weighted mean, $\tcalt$ is the vector of theoretical traveltimes $\tcal$ minus their weighted mean, where the weights $w_{i}$ are given by $w_{i}=\sum_{j}w_{ij}$ and $w_{ij}=[\Sigma_s^{-1}]_{ij}$.
The picking errors are generally considered as independent so $\Sigma_s$ has non-zero terms only on its diagonal which contains the variances of the measurement errors.
The best location is thus simply the spatial position which minimizes the misfit function:

\begin{equation}
S(x,y,z)=\sum_{i=1}^{nobs}\left(\frac{\tiobst-\ticalt(x,y,z)}{\sigma_{i}}\right)^2
\end{equation}
In practice the marginal pdf for the spatial location is estimated by applying an exhaustive grid search of the parameter space (x, y, z) as implemented by many authors such as \cite{moser_hypocenter_1992} or \cite{lomax_probabilistic_2000} .
We write here the minimum misfit function of the best location in a given velocity model for the $j^{th}$ event $S_{min}(x_j,y_j,z_j,\mathbf{m})$.

Here we aim to estimate not only the best event locations but also the best 1D velocity model. This problem can be formalized as estimating the 1D velocity model that minimizes the total misfit function over a set of events:
\begin{equation}
S(\mathbf{m})=\sum_{j=1}^{nevents}S_{min}(x_j,y_j,z_j,\mathbf{m})
\end{equation}

\begin{equation}
S(\mathbf{m})=\sum_{j=1}^{nevents}\mathbf{min}\sum_{i=1}^{nobs}\left(\frac{\tijobst-\tijcalt(x,y,z,\mathbf{m})}{\sigma_{j,i}}\right)^2
\label{eq_misfit}
\end{equation}.

In practice for each combination of velocity model parameters, the whole set of events is located in this velocity model and the total misfit function summed over the 193 events estimated.
This grid search location is repeated for the whole possible combination of velocity model parameters.

The 1D velocity model is parametrized by 5 $V_p$ and 2 $V_{p}/V_{s}$ nodes (Fig. \ref{Fig2}).
Between nodes $V_p$ and  $V_{p}/V_{s}$ are linearly interpolated.
Above first node at 0 km and below last node at 8 km, all values are set constant to nearest node.
The velocity model has been extended 1 km above sea-level to account for station elevations.
This smoother parametrization compared to the blocky velocity model should avoid a concentration of events along the interfaces.
The uniform prior information for the velocity model grid search is also represented on Fig. \ref{Fig2} where the black dots correspond to the minimum and maximum values at each of the $V_p$ and $V_{p}/V_{s}$ nodes.
The only constraint that we imposed is that the $V_p$ velocity should increase with depth.
As mentioned in the previous section, the traveltime picking accuracy of the $S$ waves was estimated in a range of 0.02 to 0.07 s, thus we expect less accurate resolution in retrieving a $V_{p}/V_{s}$ ratio and therefore decided to parametrize this profile with only 2 nodes.
We discretize the range between these limits with a step of $100-m/s$ for $V_p$ nodes and a step of 0.02 for $V_{p}/V_{s}$ nodes leading to approximately $2.10^6$ explored velocity models.

In practice for locating the events and computing the misfit function the 1D $V_p$ and $V_{p}/V_{s}$ profiles are extended to 3D $V_p$ and $V_s$ velocity grids of 180 $\times$ 470 $\times$ 470 cubic cells, of side 100m for a total extension of 18 km $\times$ 47km $\times$ 47 km.
To generate accurate $P$ and $S$ traveltimes in an efficient way we used an Eikonal solver \citep{noble_accurate_2014} that combines a spherical approximation close to the source and a plane wave approximation far away.

\begin{figure}
\centering
\includegraphics[width=11cm]{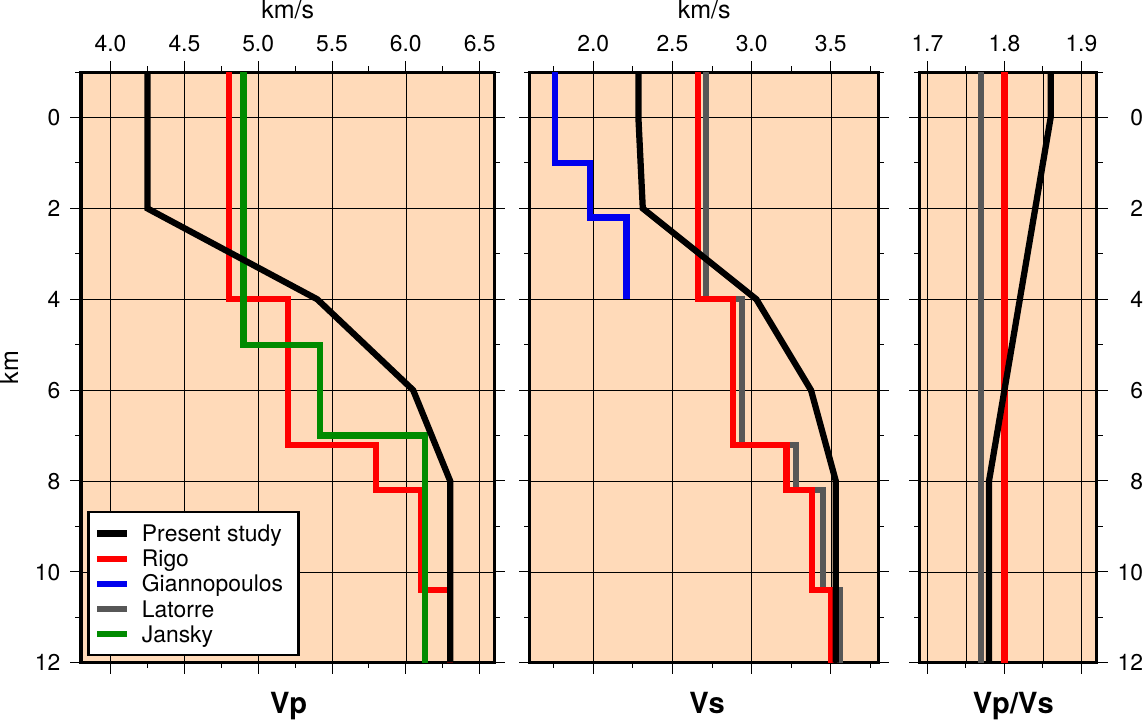}
\caption{
Left: $V_p$ profile of the best velocity model after exploring the model space (black solid line), red line $V_p$ velocity model of \cite{rigo_microseismic_1996} and green line $V_p$ velocity model proposed by \cite{jansky_estimation_2007}.
Middle: $V_s$ profile of the best velocity model after exploring the model space (black solid line), red line $V_s$ velocity model of \cite{rigo_microseismic_1996}, gray line $V_s$ velocity model of \cite{latorre_new_2004} and blue line $V_s$ model proposed by \cite{giannopoulos_ambient_2017}.
Right $V_p/V_s$ profile of the best velocity model after exploring the model space (black solid line), red line $V_p/V_s$ ratio of \cite{rigo_microseismic_1996} and gray line $V_p/V_s$ suggested by \cite{latorre_new_2004}.
}
\label{Fig4}
\end{figure}

\begin{table}
\caption{Parameters of new velocity model}
\label{Tab1}
\begin{tabular}{rccccc}
\end{tabular}
\begin{tabular}{@{}rccccc}
\hline \hline
Layer & Depth (km) & VpTop (km/s) & VpGrad (km/s/km) & VsTop (km/s) & Vsgrad (km/s/km) \\
\hline \hline
1 & -1 & 4.25 & 0 & 2.28 & 0.01 \\
2 & 2 & 4.25 & 0.575 & 2.31 & 0.36 \\
3 & 4 & 5.4 & 0.325 & 3.03 & 0.17 \\
4 & 6 & 6.05 & 0.125 & 3.37 & 0.08 \\
5 & 8 & 6.3 & 0 & 3.53 & 0 \\
6 & 15 & 6.5 & 0 & 3.67 & 0 \\
7 & 30 & 7.0 & 0 & 3.96 & 0 \\
\hline
\end{tabular}

\medskip
New velocity model described by 7 layers: depth corresponds to top of layer ($km$), $V_p$ at top of layer ($km/s$), $V_p$ velocity gradient ($km/s/km$), $V_s$ at top of layer ($km/s$), $V_s$ velocity gradient ($km/s/km$). The maximum depth of the seismic events used for this tomography being around 10-11 km we added at the bottom two extra layers as proposed by \cite{rigo_microseismic_1996}. Taking into account these extra layers might be necessary when using faraway stations to locate events in the Rift of Corinth. For these last layers we set the $V_p/V_s$ ratio to 1.78.
\end{table}

\section{Results}
\begin{figure}
\centering
\includegraphics[width=14cm]{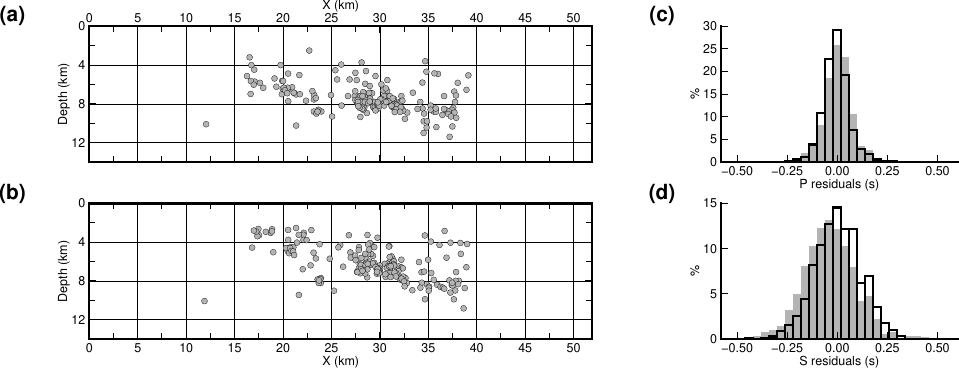}
\caption{
(a) South-North vertical section of all 193 events located in the Rigo velocity model using our non-linear grid search algorithm.
(b) South-North vertical section of the same events located in the new velocity model derived from this study.
(c) $P$ traveltime residuals: gray histogram corresponds to residuals for the Rigo velocity model, black line histogram corresponds to residuals for new velocity model.
(d) $S$ traveltime residuals: gray histogram corresponds to residuals for the Rigo velocity model, black line histogram corresponds to residuals for new velocity model.
}
\label{Fig5}
\end{figure}

After exploring the entire model space, our best $V_p$ and $V_s$ model (with the smallest misfit as defined in equation \ref{eq_misfit}) is displayed in Fig. \ref{Fig4} (black solid lines). The parameters of this model (depth of layers, $V_p$ and $V_s$ velocity at top of layers and velocity gradients) are given in table \ref{Tab1}.  
This new velocity model gives a global decrease of 30 $\%$ of the misfit function when compared to the model proposed by \cite{rigo_microseismic_1996} for the 193 events that were used for this study.
Hypocentre in this new velocity model show an overall improvement of location errors expressed as an average RMS reduction of 12 $\%$ for the P phases and 15$\%$ for the S phases.
Rigo's velocity model provides hypocentre locations with an average RMS of 0.069 s and 0.133 s for P and S phases respectively, with this new model average RMS for all events drops down to 0.061 s for the P and 0.110 s for the S phases.  
This misfit reduction is illustrated by the $P$ and $S$ histograms (Fig. \ref{Fig5}) that display the distribution of the $P$ and $S$ wave residual times for the velocity model derived from this study and the model proposed by  \cite{rigo_microseismic_1996}.
The histogram of $S$ residuals obtained with Rigo's is not centered at 0, but shifted around -0.05 s.
This negative shift corresponds to observed traveltimes smaller than predicted and thus indicating a a too high ${V_p}/{V_s}$ ratio.
Locating events in the new velocity model produces an histogram for $S$ residuals well centred at 0 s.

The main features of this new model (Fig. \ref{Fig4}) that we would like to point out is the significant decrease of $V_p$ and $V_s$ velocities from surface down to a depth of approximately 3 km compared to Rigo's model.
The $V_p/V_s$ ratio at the surface is 1.86 and decreases down to 1.78 at a depth of 8 km.
Focussing on the $S$ velocity of 2.3 km/s of the first 2 to 3 km in depth of our velocity model (in spite of a significant decrease compared to Rigo's model) does not match the average shear velocity of 2 km/s of \cite{giannopoulos_ambient_2017} obtained from ambient-noise interferometry tomography.
This difference can be explained by the fact that the tomography proposed by Giannopoulos is centred in the middle of rift and thus essentially samples soft sediments.
On the contrary our approach based on the seismicity is mostly sampling the subsurface directly below the stations that are distributed on the edges of the rift.

\begin{figure}
\centering
\includegraphics[width=16cm]{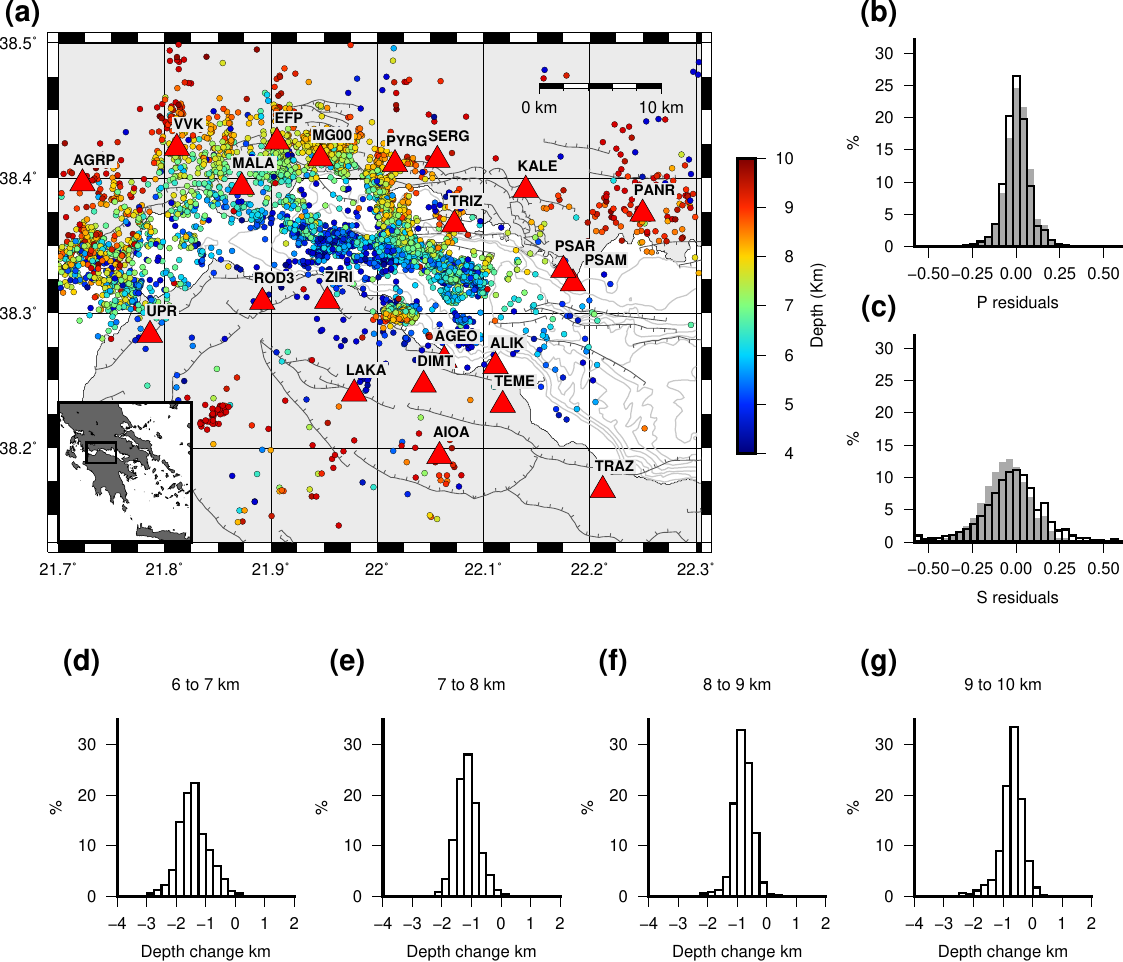}
\caption{
(a) 7400 events (same events as Fig. \ref{Fig1}) located in new velocity model using $P$ and $S$ phases (automatic picking).
For these events with automatic picks a global misfit decreases of 18 $\%$ is observed compared to Rigo' velocity model.
Average RMS decreases: 11 $\%$.
(b) $P$ traveltime residuals: gray histogram corresponds to residuals for the Rigo velocity model, black line histogram corresponds to residuals for new velocity model.
(c) $S$ traveltime residuals: gray histogram corresponds to residuals for the Rigo velocity model, black line histogram corresponds to residuals for new velocity model.
(d) to (g) histograms of focal depth variations of events for different depth ranges.
Depth change = $Z$ New model - $Z$ Rigo model.
}
\label{Fig6}
\end{figure}

A first insight of the influence of this new model on absolute event location can be observed on the South-North vertical sections (Fig. \ref{Fig5}).
These sections that are displayed with a one to one scale tend to show that longitude and latitude changes between Rigo's model and our new model for all 193 events  are fairly minor compared to depth variation.
One can first notice that many events using Rigo' blocky velocity model are concentrate around the different layer interfaces with strong velocity contrasts at 4 and 7.2 km.
Our smoother parametrization avoids such artefacts.
On the whole, the depth of events tend to decrease when using the new model.
The depth of hypocentres deceases by several hundreds of metres for events located around 11 to 8 km in depth. The decrease in depth becomes greater for shallower events.
The shallowest events selected with an average depth around 4 km move upwards to the surface by more than 1 km.

To further assess the influence of this new 1D velocity profile on event locations, we relocated all 7400 earthquakes previously presented in Fig. \ref{Fig1}. Results are displayed in Fig. \ref{Fig6}.
For this dataset we are using $P$ and $S$ phases that were inferred using an automatic picker and we observe a global decrease of 18 $\%$ of the misfit function when compared to the model proposed by Rigo.
This 18 $\%$ corresponds to an average RMS reduction of about 8 $\%$ for both P and S phases.
The histograms of Fig. \ref{Fig6} showing  focal depth variations of the 7400 events for different depth ranges confirm the fact that depth of events decrease using the model that we derived during this study.
Depth change varies in average from 500 meters for events located in depth (8 to 11 km) to 1.5 km for events close to the surface.

\section{Discussion \& Conclusion}

The objective of this study is to propose a new updated accurate 1-D $V_p$ and $V_s$ velocity model of the western Rift of Corinth for precise absolute earthquake locations.
The methodology used to obtain this new model associates two fully non-linear algorithms, a complete exploration of the $V_p$ and $V_s$ model space combined with a grid search method for earthquake locations to minimize the P and S arrival time residuals.
To try and constrain as much as possible this inverse problem, we first selected seismic events with in average a minimum of 20 $P$ phases and a minimum of 15 $S$ phases.
Concerning the depth of the selected events, depth ranges from approximately 4 km down to 10 km, with only a few events between 10 and 12 km depth.
We calculated the misfit function for approximately $2.10^6$ velocity models.
For optimal model we observe  global decrease of 30 $\%$ of the misfit function when compared to the model proposed by Rigo.
This misfit function decrease corresponds to an overall improvement of location errors expressed in terms of RMS to an average reduction of about 14 $\%$.
The major influence on the locations of events is a global decrease of focal depths that range from a few hundred metres for deep events to more than one kilometre for shallow earthquakes.

\begin{figure}
\centering
\includegraphics[width=13cm]{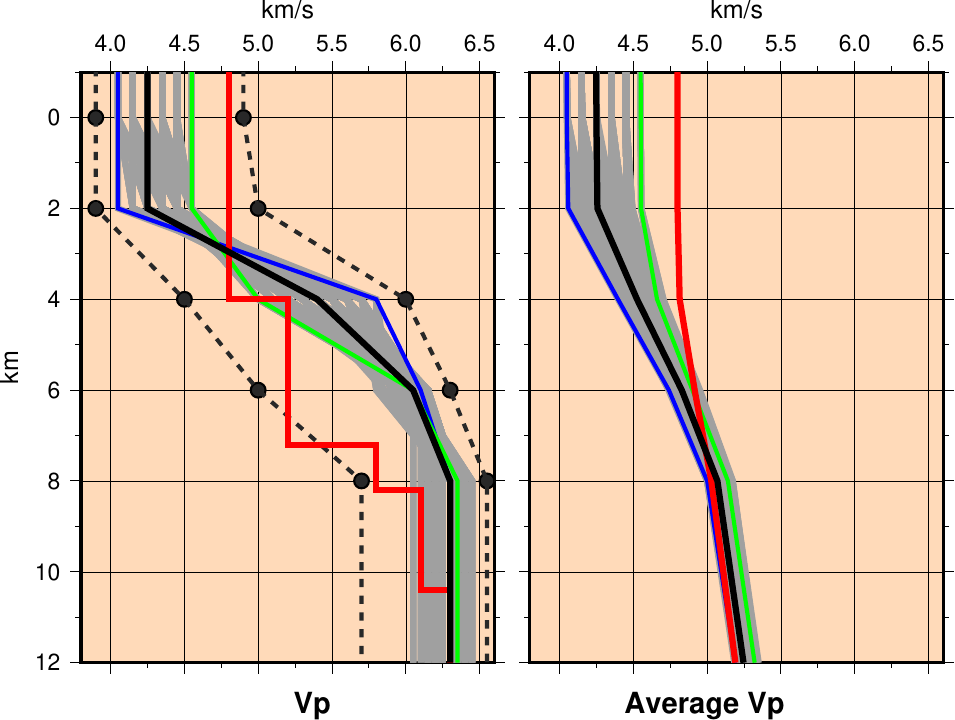}
\caption{
Right: $V_p$ velocity profiles.
Left: $V_p$ Average velocity (for each point in depth average velocity from surface).
Black lines: $V_p$ best model with a misfit reduction 30 $\%$ compared to Rigo’s model.
Red line $V_p$ Rigo model.
Gray lines correspond to all $V_p$ velocity profiles with a misfit decrease greater than 25 $\%$ compared to Rigo’s model.
}
\label{Fig7}
\end{figure}

After exploring the $2.10^6$ models, we did not identify any secondary minimum, we found only a single  global minimum.
Displaying the histogram of the 1 D marginal misfit for each individual inverted parameter (velocity nodes) along with its associated confidence interval is not really the best criterion to assess the accuracy / uncertainty of our new velocity model.
The main reason is that there a strong coupling between some parameters.
To illustrate this coupling between parameters, we displayed in Fig. \ref{Fig7} all the $V_p$ velocity profiles with a misfit decrease greater than 25 $\%$ compared to Rigo’s model.
Two extreme velocity profiles (velocity function of depth) that are highlighted (green and blue lines) show very different velocities in the depth range 0 to 4 km.
It turns out that if we convert all these velocity models to average velocities (for each point in depth average velocity between current depth and surface) all models are cinematically equivalent to the best model (black line) within a maximum range of $\pm 115 m/s$ for depths of 4 km and onwards.
For the depth range 0 to 2 - 3 km, average velocities for these models (that all have a very good misfit compared to Rigo's model) span over a wider window $\pm 250 m/s$.
This larger dispersion of average velocities close to the surface tend to indicate that inverted velocities are probably less constrained due to the lack of seismic events in this depth range (0 and 4 km).
All the average velocity profiles displayed in Fig. \ref{Fig7} show lower velocities down to a depth of 8 km compared to Rigo's model.

In this paper, we improved the 1D P and S velocity models for the Western rift of Corinth, and this velocity model should be used in the future for absolute seismic event locations.
In this model focal depths of events compared to Rigo’s model vary form a few hundred meters to more than 1 km as we come close to surface and will certainly modify the geometry (depth and dip) of the seismically actives structures that have already been mapped. 
The next natural steps should be to relocate all the events of CRL catalogues and also to invert for the 3D velocity structure using the model as initial guess. 
 
\begin{acknowledgments}
The authors thank the personnel of Corinth Rift Laboratory network (CRLnet) and Hellenic Unified Seismological Network (HUSN) who worked for the installation, operation, and maintenance of stations used in this article.
The figures presented in this paper were created using the Generic Mapping Tool software \citep{wessel_new_1998}.

\end{acknowledgments}

\section{Data availability}

Seismic data used in this study can be retrieved from the CRL net-work (Corinth Rift Laboratory, http://crlab.eu) for stations under the code CL. For stations under the code HA and HP the data can be retrieved from the Hellenic Unified Seismological Network (HUSN, \\ http://www.gein.noa.gr/en/networks/husn).

\bibliography{Corinth-1}

\label{lastpage}

\end{document}